\documentclass[amsmath,amssymb,superscriptaddress,twocolumn, showkeys, prl]{revtex4-1}

\usepackage{graphicx}% Include figure files
\usepackage{dcolumn}% Align table columns on decimal point
\usepackage{bm}% bold math
\usepackage{color}
\begin{document}

\title{Chemical sensing by nonequilibrium cooperative receptors}

\author{Monica Skoge} \affiliation{Department of Biology, University
  of California San Diego, La Jolla, CA 92093-0319}

\author{Sahin Naqvi}
\altaffiliation{Present address:  Department of Biology, MIT}
\affiliation{Department of Molecular Biology, Princeton University,
Princeton, NJ 08544}

\author{Yigal Meir}
\affiliation{Department of Physics, Ben-Gurion University,
Beer Sheva, Israel 84105}

\author{Ned S. Wingreen}
\affiliation{Department of Molecular Biology, Princeton University,
Princeton, NJ 08544}

\begin{abstract}
Cooperativity arising from local interactions in equilibrium receptor
systems provides gain, but does not increase sensory performance, as
measured by the signal-to-noise ratio (SNR) due to a fundamental
tradeoff between gain and intrinsic noise. Here we allow sensing to be
a nonequilibrium process and show that energy dissipation cannot
circumvent the fundamental tradeoff, so that SNR is still optimal for
independent receptors.  For systems requiring high gain,
nonequilibrium 2D-coupled receptors maximize SNR, revealing a new
design principle for biological sensors.
\end{abstract}

\maketitle

Biological systems generally operate out of equilibrium, using
free-energy dissipation to drive metabolic reactions, perform
mechanical work via molecular motors, communicate over large distances
via action potentials, and much more.  In cellular information
processing, free-energy dissipation plays essential roles in
adaptation~\cite{Yuhai2012}, time-averaging~\cite{Mehta2012}, and the
sensitivity of chemical switches~\cite{Qian2005}, and can overcome
equilibrium physical limits to
performance~\cite{kineticproofreading1974, Yuhai2008}.  The best-known
example of the latter is kinetic
proofreading~\cite{kineticproofreading1974}, in which the specificity
of interaction between an enzyme and two competing substrates can
exceed the equilibrium limit set by the ratio of their binding
affinities.  Previously, we identified a tradeoff between gain and
intrinsic noise for equilibrium {\em locally} coupled receptor
systems, limiting their ability to sense weak signals, as measured by
the signal-to-noise ratio (SNR)~\cite{skogeprl2011}.  Can free-energy
dissipation also help circumvent this tradeoff and thereby increase
sensory performance?

In what follows, we answer this question generally for 1D- and
2D-coupled receptor systems by optimizing the SNR over the full,
nonequilibrium parameter space of these systems, subject only to
constraints of lattice symmetry and locality of interactions.
Compared to equilibrium, these systems gain additional parameters
related to cyclic fluxes, which enable novel
behavior. Our main results are: (a) even for
  nonequilibrium the SNR is optimal for independent receptors
  compared to coupled receptors, and (b) when gain (and hence
  cooperativity) is required, nonequilibrium can improve SNR for
  2D-coupled receptors, but not for 1D-coupled receptors. The first
  result is an extension of our previous equilibrium observations,
  showing that nonequilibrium can at best modestly alter the tradeoff between
  gain and intrinsic noise due to interaction-mediated slowing down.  To
  understand the second, surprising result we map the dynamics of
  cooperative receptors onto simpler 1-step processes, and uncover an
  optimal design principle for biological sensors with high gain.

A classic problem facing the cell, originally posed by Berg and
Purcell~\cite{BergPurcell1977}, is the estimation of external ligand
concentration via cell-surface receptors in the presence of stochastic
fluctuations.  Here, we consider a variant of this problem motivated
by {\it E. coli}'s strategy of sensing small {\em changes} in
chemotactic ligand concentration using strongly-coupled
chemoreceptors. Following our previous framework~\cite{skogeprl2011},
we study the linear response of the average total activity $A \in [0,
  N]$ of $N$ coupled receptors to a small relative change in ligand
concentration $\Delta \log(\text{[L]})$, time-averaged over a period
$\tau_{\text{avg}}$ ({\it e.g.} set by the turnover time of the
response regulator CheY-P in {\it E. coli}) that is long.  The sensing
performance of the cell is governed by the signal-to-noise ratio per
receptor~\cite{skogeprl2011}
\begin{equation}
\text{SNR}(\tau_{\text{avg}}) \equiv \frac{(\Delta A)^2}{N
  \sigma^2(\tau_{\text{avg}})},\label{SNR}
\end{equation}
where $\Delta A\sim\Delta \log(\text{[L]})$ is the resulting change in
average activity and $\sigma^2(\tau_{\text{avg}})$ is the
long-time-averaged variance of activity, which decreases as
$\sim1/\tau_{\text{avg}}$ as $\tau_{\text{avg}}\rightarrow \infty$.

We model receptor cooperativity using a general Ising-type framework
with {\em local} receptor-receptor interactions.  By varying the
coupling strength $J$, our model encompasses independent receptors
($J=0$), receptors near a critical point (intermediate $J\approx
J_c$), as well as allosteric Monod-Wyman-Changeaux (MWC)
receptors~\cite{Monod1965} (large $J$).  This last regime has had much
success explaining steady-state signal amplification by bacterial
chemoreceptors.  Importantly, finite $J$ intrinsically slows down the
rate of receptor switching, thereby limiting noise reduction via
time-averaging~\cite{skogeprl2011}.  Indeed, this slowing down is most
dramatic in the strongly-coupled (large $J$) limit and corresponds to
the very slow switching of a ferromagnet below the transition
temperature -- a process generally considered too slow to be relevant
for magnets, but required for signaling by MWC systems.  The previous
successful applications of the MWC model to bacterial chemoreceptors
motivate our question whether nonequilibrium can counteract the
slowing down of receptor switching and increase sensory performance.

\begin{figure}
\begin{center}
\includegraphics*[scale=0.2,angle=0]{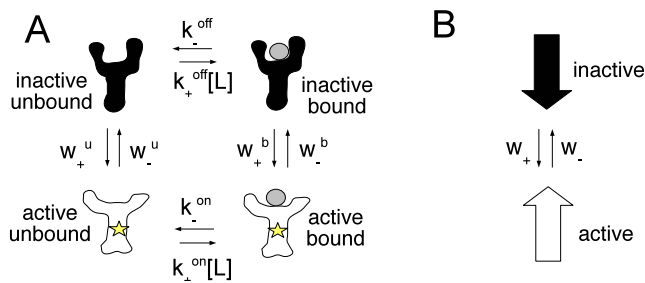}
\caption{\textbf{Single receptor.}  \textbf{(A)} Schematic diagram of
  interconversion among the four states of a single receptor:
  inactive/unbound, inactive/bound, active/bound, and active/unbound.
  \textbf{(B)} In the limit of fast ligand binding and unbinding, the
  receptor dynamics reduces to switching between two states, active
  and inactive.  }
\label{ligandstates}
\end{center}
\end{figure}

Specifically, we consider four-state receptors, which can bind/unbind
ligand with rates $\{k\}$ and switch conformations between active and
inactive states with rates $\{w\}$, as shown schematically in
Fig.~\ref{ligandstates}(A).  As allosteric coupling between receptors
is mediated via the conformational degrees-of-freedom as opposed to
the occupancy of binding pockets, we let the switching rates $\{w\}$
depend on the activity (but not the occupancy) of
neighboring receptors, while keeping the binding/unbinding rates
$\{k\}$ independent of neighboring receptors.
Conformational switching rates $\{w\}$ ($\leq 10^4/$s for large-scale
rearrangements~\cite{SmockGierasch2009}) are typically slower than the
binding/unbinding rates $\{k\}$ ($\approx 10^5/$s for bacterial
chemoreceptors~\footnote{Assuming diffusion-limited
  binding~\cite{BergPurcell1977} with diffusion coefficient $D=400
  ~\mu \text{m}^2/s$ and receptor size $s = 10$~nm gives an
  association rate $k_+ = 4Ds = 3.9e9/\text{M}s$.  For observed
  chemotactic ligand affinities of $\approx
  20~\mu\text{M}$~\cite{Keymer2006}, the corresponding dissociation
  rates is $k_- = 7.7e4/s$.  }), and, therefore, many individual
binding/unbinding events contribute to an effective field $f$ that
biases receptor activity via the differences in ligand affinity
between the active and inactive states.  In this fast binding limit,
the dynamics of four-state receptors can be reduced to that of
effective two-state receptors, as shown schematically in
Fig.~\ref{ligandstates}(B), with changes in ligand concentration
implying changes in the field $f$.  Consequently, we can write the
signal $\Delta A$ in Eq.~(\ref{SNR}) as
\begin{equation}
\Delta A=\frac{1}{4}R \Delta f,
\end{equation}
where we define the response $R \equiv 4dA/df$ and the gain $R/N \in
[1, N]$ is the amplification of changes in activity relative to
independent receptors due to receptor-receptor interactions.  In what
follows, we deal exclusively with coupled two-state receptors and
optimize SNR/[$\tau_{\text{avg}}(\Delta f)^2$] to focus on the
benefits of nonequilibrium for receptor cooperativity~\footnote{By
  restricting our analysis to coupled two-state receptors rather than
  four-state receptors, we neglect possible nonequilibrium coupling
  between ligand binding and receptor interactions. }.  (For a full
discussion of the role of nonequilibrium in determining the
sensitivity of the four-state receptor, see~\cite{suppinfo}.)

The behavior of the coupled two-state receptor system is completely
determined by the conformational switching rates $\{w\}$, and thus we
seek to optimize these rates to maximize SNR, without imposing the
equilibrium constraint of detailed balance.  As a simple
multiplicative increase of all rates can trivially decrease the
correlation time, thereby decreasing noise and increasing SNR, we
constrain the sum of forward and backward rates for conformational
switching to be $w_++w_-=\alpha$, where $\alpha$ is the intrinsic
switching rate (with units of inverse time)~\footnote{We investigated
  letting $w_++w_-\leq \alpha$, but found SNR was always optimal for
  equality. }.  In this case, the rates $\{w\}$ can be expressed in
the form of heat-bath kinetics
\begin{equation}
w_{\pm} = \frac{\alpha}{1+e^{\pm \Delta \epsilon}},\label{switchingrates}
\end{equation}
where $\Delta \epsilon$ is an effective energy change upon switching
conformation from inactive to active~\footnote{We work in units such
  that $k_BT=1$, $\alpha=1$ and $K^{\text{off}}=1$.}.

\begin{figure}
\begin{center}
\includegraphics*[scale=0.15,angle=0]{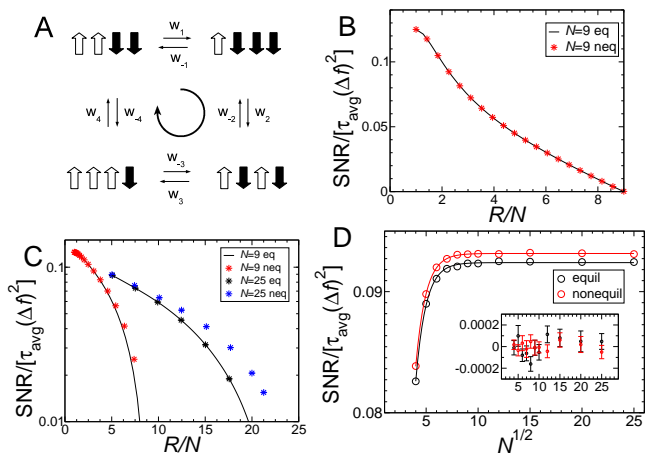}
\includegraphics*[scale=0.16,angle=0]{figure2b.eps}
\includegraphics*[scale=0.16,angle=0]{figure2c.eps}
\includegraphics*[scale=0.16,angle=0]{figure2d.eps}
\caption{\textbf{1D- and 2D-coupled receptors.}  \textbf{(A)}
  Schematic diagram of 4-state cycle for a chain of 1D-coupled
  receptors. White-up and black-down arrows denote active and inactive
  receptors, respectively.  Optimal SNR/[$\tau_{\text{avg}}(\Delta
    f)^2$] as a function of gain $R/N$ for \textbf{(B)} 1D- and
  \textbf{(C)} 2D-coupled receptors for equilibrium (black) and
  nonequilibrium (red $N=9$ and blue $N=25$).  \textbf{(D)}
  SNR/[$\tau_{\text{avg}}(\Delta f)^2$] as a function of linear system
  size $N^{1/2}$ for gain $R/N=5$ for equilibrium (black) and
  nonequilibrium with $\gamma_J=-0.5$ (red).  Error bars show standard
  errors of the mean and curves are exponential fits.  The inset shows
  the residuals from the fits. }
\label{nonequil2Dfig}
\end{center}
\end{figure}

{\it 1D receptors.} We first consider a 1D chain of $N$ receptors. For
the 1D-chain, there is only one nonequilibrium degree-of-freedom
$\gamma_{f2} \equiv \Delta G/2$, which we define in terms of the
thermodynamic driving force $\Delta G$ of the 4-cycle shown in
Fig.~\ref{nonequil2Dfig}(A): For any cycle in a nonequilibrium
steady-state system, the thermodynamic driving force $\Delta G$ is
related to the ratio of the cycle fluxes, $j_{\text{CW}}/j_{\text{CCW}}$, in
the clockwise versus counterclockwise directions, or equivalently, the
ratio of the product of rate constants going clockwise around the loop
to that going counterclockwise, according to~\cite{hillbook,
  BeardQian2007}
\begin{equation}
e^{-\Delta G} = \frac{j_{\text{CW}}}{j_{\text{CCW}}} = \frac{w_{1} w_{2}
  w_{3}w_{4}}{w_{-1}w_{-2} w_{-3} w_{-4}}, 
\end{equation}
In equilibrium, these products are equal so that $\Delta G = 0$.
Therefore, the $\Delta G$'s for cycles in reaction space provide a
useful basis for parameterization of the nonequilibrium behavior.  The
energy dissipation rate attributable to a given cycle is the product
of $\Delta G$ and the net cycle flux $j_{\text{CW}} - j_{\text{CCW}}$ (in analogy to
power being the product of voltage and current in electrical
circuits~\cite{hillbook}).

The switching rates $\{w\}$ from Eq.~(\ref{switchingrates}) are
parameterized in Table~\ref{glauberenergies} in terms of $\gamma_{f2}$
and the equilibrium Ising parameters~\cite{Glauber1963}: the coupling
strength $J$ and the field $f$.  The nonequilibrium parameter
$\gamma_{f2}$ behaves as an additional field present only when nearest
neighbors match each other's activity.  As expected, the model reduces
in equilibrium ($\gamma_{f2}=0$) to the 1D Ising model with heat bath
kinetics~\cite{Glauber1963}.
\begin{table}
\begin{center}
\begin{tabular}{|c|c|c|}
\hline
1D Neighbors & $\Delta \epsilon$ \\
\hline
\hline
$++$ & $-4J + f + \gamma_{f2} $ \\
$-+$ & $f$ \\
$--$ & $4J + f + \gamma_{f2} $ \\
\hline
\hline
2D Neighbors & $\Delta \epsilon$ \\
\hline
\hline
$++++$ & $-8J + f + \gamma_{f4} $ \\
$-+++$ & $-4J - \gamma_J + f + \gamma_{f3}$ \\
$--++$ & $f$ \\
$---+$ & $4J + \gamma_J + f + \gamma_{f3}$ \\
$----$ & $8J + f + \gamma_{f4} $ \\
\hline
\end{tabular}
\end{center}
\caption{Energy change $\Delta \epsilon$ upon switching conformation
  from inactive to active with a given configuration of nearest
  neighbors. }
\label{glauberenergies}
\end{table}  

For a 1D-chain with $N=9$, we globally searched the $J,f,\gamma_{f2}$
parameter space to maximize the SNR numerically utilizing a
nonequilibrium fluctuation-dissipation
theorem~\cite{noneqFDT2009,suppinfo}.  We found that SNR is globally
optimal for independent receptors, as shown in
Fig.~\ref{nonequil2Dfig}(B).  Moreover, even for a specified gain,
nonequilibrium offers no advantage.

{\it 2D receptors.} Can nonequilibrium improve SNR in higher
dimensions?  For a 2D square lattice with nearest-neighbor
interactions and four-fold symmetry~\footnote{For simplicity, we
  assume the same switching rates for the states with neighbors $++--$
  or $+-+-$ around the cardinal directions.}, there are three
nonequilibrium degrees-of-freedom, which we define as $\gamma_{f3}$,
$\gamma_{f4}$, and $\gamma_J$.  As in 1D, we define these
nonequilibrium parameters in terms of the driving forces for 2D
versions of the cycle shown in
Fig.~\ref{nonequil2Dfig}(A)~\cite{suppinfo}.  The 2D switching rates
$\{w\}$ from Eq.~(\ref{switchingrates}) are given in
Table~\ref{glauberenergies}. Like $\gamma_{f2}$ of the 1D model, the
nonequilibrium parameters $\gamma_{f3}$ and $\gamma_{f4}$ behave as
additional fields specific for the states with three and four
activity-matched nearest neighbors, respectively.  The third
nonequilibrium parameter $\gamma_J$ behaves as an additional coupling
strength for states with exactly three matched nearest neighbors.

For 2D receptors, a global search of the 5-dimensional
$J,f,\gamma_{f3},\gamma_{f4},\gamma_J$ parameter space reveals, once
again, that SNR is maximal for independent equilibrium receptors.
However, unlike 1D, for a specified gain $R/N>1$, nonequilibrium does
increase the SNR relative to nonequilibrium, as shown in
Fig.~\ref{nonequil2Dfig}(C), via an ``antiferromagnetic''
pseudo-coupling $\gamma_J < 0$. Though modest, this improvement is not simply a
finite-size effect and persists in the limit $N\rightarrow \infty$, as
shown in Fig.~\ref{nonequil2Dfig}(D) with $\gamma_J=-0.5$ chosen for
convenience (since $\gamma_J$ is not optimal this provides a lower
bound on the increase in SNR achievable by nonequilibrium).

\begin{figure}
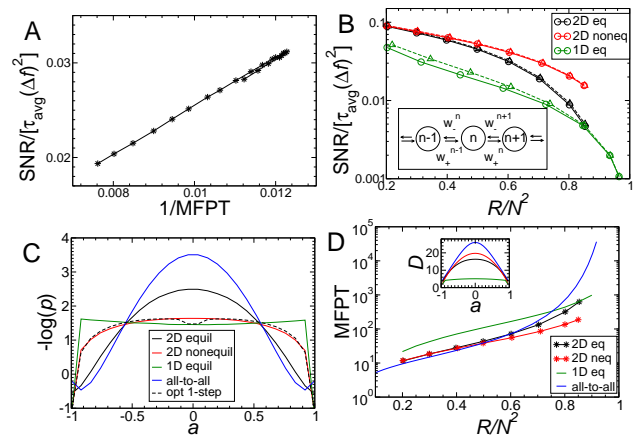

\begin{center}
\includegraphics*[scale=0.15,angle=0]{figure3a.eps}
\includegraphics*[scale=0.15,angle=0]{figure3b.eps}
\includegraphics*[scale=0.15,angle=0]{figure3c.eps}
\includegraphics*[scale=0.15,angle=0]{figure3d.eps}
\caption{\textbf{SNR, MFPT, and effective potentials at fixed gain.}
  \textbf{(A)} SNR/[$\tau_{\text{avg}}(\Delta f)^2$] versus 1/MFPT for
  2D-coupled receptors with $N=25$ for gain $R/N=0.7N$ as $\gamma_J$
  is increased from zero (starting at left).  \textbf{(B)}
  SNR/[$\tau_{\text{avg}}(\Delta f)^2$] versus normalized gain $R/N^2$
  for 1D- and 2D-coupled receptors and their mappings to 1-step
  processes (dashed lines).  Inset shows schematic of 1-step process
  with transition probabilities~\cite{vankampen}.  \textbf{(C)}
  ``Potentials'', $-\log{p(a)}$, as functions of normalized and
  symmetrized activity $a=2A/N-1$ for Ising lattices with $N=25$ for
  gain $R/N=0.7N$ and the optimal 1-step process (dashed curve).
  \textbf{(D)} MFPT as a function of $R/N^2$(=gain/$N$) for Ising
  lattices with $N=25$.  Inset shows the diffusion coefficients $D(a)$
  for gain $R/N=0.7N$.  }
\label{MFTfig}
\end{center}
\end{figure}

Why does nonequilibrium improve SNR for 2D-coupled receptors (but not
1D-coupled receptors)?  For a fixed gain $R/N$, optimizing SNR is
equivalent to minimizing the noise $\sigma^2(\tau_{\text{avg}})$.  We
find that nonequilibrium does not decrease the steady-state variance
(``snapshot noise'')~\cite{suppinfo}, and therefore it must decrease
the correlation time of the noise $\tau_c$, making time-averaging more
effective.  In the high-gain regime, the steady-state distribution of
activity $p(A)$ is strongly peaked at the fully active/inactive states
and the correlation time is approximately the mean-first-passage-time
(MFPT) to switch from fully inactive/active to half-maximal
activity. As shown in Fig.~\ref{MFTfig}(A), the SNR for
nonequilibrium optima depends linearly on the inverse MFPT as the
nonequilibrium parameter $\gamma_J$ is varied, showing that
nonequilibrium indeed improves SNR by minimizing MFPT.

{\it Activity as a reaction coordinate.} To better understand how
nonequilibrium decreases MFPT at high gain, we map the dynamics of our
coupled receptor systems with their $N$-dimensional state space onto a
single dimension, a suitable reaction coordinate, which we take to be
the normalized and symmetrized total activity $a\equiv2A/N-1$.  The
dynamics then takes the form of an analytically tractable 1-step
process~\cite{vankampen}, as shown schematically in the inset to
Fig.~\ref{MFTfig}(B).  For the forward and backward rates $w_{\pm}^n$,
we take the trajectory-averaged transition rates from simulations of
the dynamics in the full state space.  This mapping to a 1-step
process preserves the dynamics surprisingly well, specifically
capturing the SNR as a function of gain as shown in
Fig.~\ref{MFTfig}(B) for both 1D- and 2D-coupled receptors.

The dynamics of 1-step processes are fully described by the
steady-state probability distribution $p(a)$ and diffusion coefficient
$D(a) \equiv (w_+^n+w_-^{n+1})(N\Delta a)^2/2$, where $\Delta a=2/N$ is
the spacing between states along the reaction coordinate.  The
relationship between MFPT, $p(a)$, and $D(a)$ has been calculated in
Ref.~\cite{vankampen} and in the continuum $N\rightarrow \infty$ limit
is
\begin{equation}
\text{MFPT} = N^2\int_{-1}^0 da \frac{1}{D(a)p(a)}\int_{-1}^{a}da'p(a').\label{MFPT}
\end{equation}
Using Eq.~(\ref{MFPT}) we can ask what are the optimal $p(a)$ and
$D(a)$ that minimizes MFPT for a certain gain.  This question is only
meaningful if the transition rates are constrained from diverging.
For simplicity, we fix the total sum of transition rates $\sum_i
(w_+^i + w_-^i)$ or in the continuum limit the integral of $D(a)$ to
be constant.  The resulting optimal ``potential'', $-\log{p(a)}$, is
shown in Fig.~\ref{MFTfig}(C) (dashed curve) and consists of a flat
barrier separating deep wells at the extreme activity
states~\cite{suppinfo}.  The deep wells make the first transition out
of the fully active/inactive state rate limiting and ensure high gain.
Subsequently, the flat potential gives a high probability for a fast
direct switch to the opposite activity state.  This mesa-like shape of
the potential is also optimal when diffusion is held
constant~\cite{suppinfo}.

Strikingly, the optimal 1-step potential agrees almost precisely with
the shape of the 2D nonequilibrium effective potential, as shown in
Fig.~\ref{MFTfig}(C).  Comparison to the corresponding 2D equilibrium
potential shows that nonequilibrium decreases MFPT by flattening the
potential along the reaction coordinate.  This insight also offers
a simple explanation why there is no benefit of nonequilibrium in 1D:
equilibrium 1D-coupled receptors already have a flat potential (green
curve), as switching in 1D involves the unbiased diffusion of a single
domain boundary separating active and inactive receptors.

What dimensionality of coupling yields the highest SNR, {\it
  i.e.} the fastest MFPT, for a fixed gain?  In addition
to the shape of the potential, the magnitude of the diffusion
coefficient $D(a)$ is also important in determining MFPT.  As shown in
the inset to Fig.~\ref{MFTfig}(D), 1D-coupled receptors have
strikingly lower diffusion coefficient than receptors with
higher-dimensional coupling.  Diffusion along the reaction coordinate
in 1D at high gain occurs exclusively via switching of one of the two
receptors at the domain boundary.  In contrast, multiple viable
switching trajectories are possible in 2D.  In the extreme case of
all-to-all coupling, which can in principle be achieved by long-range
interactions or a rapidly diffusable factor, any of the $N$ receptors
can always switch, thereby increasing the diffusion coefficient along
the reaction coordinate.  However, all-to-all coupling yields an unfavorable
potential barrier of high curvature, as shown in Fig.~\ref{MFTfig}(C).
Consequently, nonequilibrium 2D-coupled receptors achieve the fastest
switching in the high gain regime, as shown in Fig.~\ref{MFTfig}(D),
because they have the combined benefits of a relatively flat effective
potential and a large diffusion coefficient.  

To summarize, nonequilibrium cannot solve the problem of
interaction-mediated slowing down inherent in cooperativity achieved
via local receptor-receptor interactions.  Hence, independent
receptors maximize signal-to-noise ratio (SNR).  However,
nonequilibrium can increase SNR for a fixed gain by minimizing the
mean-first-passage time (MFPT) of switching between extreme activity
states.  More generally, optimizing the tradeoff between cooperativity
and interaction-mediated slowing down at fixed gain leads to a novel
optimal-design principle for chemical sensing networks -- namely
bistable effective potentials with flat barriers separating deep
wells.  For sensing via coupled-receptor systems, we have found an
unexpected near-optimality of nonequilibrium 2D-coupled receptors,
which may offer insight into the organization of {\it E. coli}
chemoreceptors.

\begin{acknowledgements}
This work was supported in part by National Science Foundation Grant
PHY-0957573 and by the National Institutes of Health (www.nih.gov)
Grant R01 GM082938.
\end{acknowledgements}

\bibliography{bibliography}

\end{document}